\documentclass[fleqn,usenatbib]{mnras}

\usepackage{graphicx}	% Including figure files
\usepackage{amsmath}	% Advanced maths commands
\usepackage{amssymb}	% Extra maths symbols
\usepackage{threeparttable} % for notes in table

\usepackage[T1]{fontenc}
\usepackage{ae,aecompl}

\usepackage{newtxtext,newtxmath}

\graphicspath{{plots/}}
\newcommand{\bc}{\begin{center}}
\newcommand{\ec}{\end{center}}

\newcommand{\hMsun}{~h^{-1}\>{\rm M_\odot}}
\newcommand{\Mpc}{~h^{-1}~{\rm Mpc}}

\newcommand{\Hunit}{~h~{\rm km}~s^{-1}~{\rm Mpc}^{-1}}

\defcitealias{Cui2018}{Paper~I}
\title[The evolution of the LSE]{The large-scale environment from cosmological simulations II: The redshift evolution and distributions of baryons}

\author[Weiguang Cui et al.]
{\parbox{\textwidth}{Weiguang Cui,$^{1,2}$\thanks{E-mail: \texttt{wcui@roe.ac.uk}}
  Alexander Knebe,$^{1,3,4}$ Noam I. Libeskind,$^{5,6}$ Susana Planelles,$^{7}$ Xiaohu Yang,$^{8,9}$ Wei Cui,$^{10}$ Romeel Dav\'{e},$^{2,11}$ Xi Kang,$^{12}$ Robert Mostoghiu,$^1$ Lister Staveley-Smith,$^{4,13}$ Huiyuan Wang,$^{14,15}$, Peng Wang,$^{5,12}$ Gustavo Yepes$^{1,3}$.
 }\vspace{0.4cm}
 \\
 \parbox{\textwidth}{
  $^1$Departamento de F\'isica Te\'{o}rica, M\'{o}dulo 15, Facultad de Ciencias,
  Universidad Aut\'{o}noma de Madrid, 28049 Madrid, Spain\\
  $^2$Institute for Astronomy, University of Edinburgh, Royal Observatory, Edinburgh EH9 3HJ, United Kingdom\\
  $^3$Centro de Investigaci\'{o}n Avanzada en F\'isica Fundamental (CIAFF), Facultad de Ciencias, Universidad Aut\'{o}noma de Madrid, 28049 Madrid, Spain\\
  $^{4}$International Centre for Radio Astronomy Research (ICRAR), University of Western Australia, 35 Stirling Highway,
  Crawley, Western Australia 6009, Australia\\
  $^5$Leibniz-Institut f\"ur Astrophysik Potsdam (AIP), An der Sternwarte 16, D-14482 Potsdam, Germany\\
  $^6$l'Institut de Physique Nucl\'{e}aire de Lyon (IPNL), University of Lyon; UCB Lyon 1/CNRS/IN2P3; Lyon, France. \\
  $^{7}$Departament d'Astronomia i Astrof{\'i}sica, Universitat de Val\`encia, c/ Dr. Moliner, 50, 46100 - Burjassot (Valencia), Spain\\
  $^8$Department of Astronomy, Shanghai Key Laboratory for
  Particle Physics and Cosmology, Shanghai Jiao Tong University,
  Shanghai 200240, China\\
  $^9$IFSA Collaborative Innovation Center, and Tsung-Dao
  Lee Institute, Shanghai Jiao Tong University, Shanghai 200240,
  China\\
  $^{10}$Tsinghua Center for Astrophysics, Department of Physics, Tsinghua University, Beijing 100084, China.\\
  $^{11}$University of the Western Cape, Bellville, Cape Town 7535, South Africa\\
  $^{12}$Purple Mountain Observatory, the Partner Group of MPI f\"ur Astronomie, No. 8 Yuan Hua Road, 210034 Nanjing, China\\
  $^{13}$ARC Centre of Excellence for All-Sky Astrophysics (CAASTRO)\\
  $^{14}$Key Laboratory for Research in Galaxies and Cosmology, Department of Astronomy, University of Science and Technology of China, Hefei, Anhui 230026, China\\
  $^{15}$School of Astronomy and Space Science, University of Science and Technology of China, Hefei 230026, China\\
}}

\date{Accepted XXX. Received YYY; in original form ZZZ}

\pubyear{2019}

\begin{document}
\label{firstpage}
\pagerange{\pageref{firstpage}--\pageref{lastpage}}
\maketitle

\begin{abstract}
Following \cite{Cui2018} (hereafter Paper I) on the classification of large-scale environments (LSE) at $z$ = 0, we push our analysis to higher redshifts and study the evolution of LSE and the baryon distributions in them. Our aim is to investigate how baryons affect the LSE as a function of redshift. In agreement with \citetalias{Cui2018}, the baryon models have negligible effect on the LSE over all investigated redshifts. We further validate the conclusion obtained in \citetalias{Cui2018} that the gas web is an unbiased tracer of total matter -- even better at high redshifts. By separating the gas mainly by temperature, we find that about 40 per cent of gas is in the so-called warm-hot intergalactic medium (WHIM). This fraction of gas mass in the WHIM decreases with redshift, especially from $z = 1$ (29 per cent) to $z = 2.1$ (10 per cent). By separating the whole WHIM gas mass into the four large-scale environments (i.e. voids, sheets, filaments, and knots), we find that about half of the WHIM gas is located in filaments. Although the total gas mass in WHIM decreases with redshift, the WHIM mass fractions in the different LSE seem unchanged.
\end{abstract}


%*******************************************************************************

\section{Introduction}
\label{i}

In the early Universe, matter is thought to have been distributed almost homogeneously, with very small fluctuations in the density field. These ``initial condtions'' are primarily evidenced by the smoothness of the CMB and the power spectrum of its inhomogenities amounting to roughly one part in 10$^5$. Due to the nature of gravitational instability, small over-densities in the matter distribution are able to overcome the cosmological expansion and collapse to form potential wells, that may virialize into so-called dark matter halos \citep[see][for example]{White1978}. If the matter budget of the Universe is dominated by Cold Dark Matter (CDM), then smaller dark matter halos attract each other and merge together to form larger halos, growing according to a hierarchical scenario \citep[e.g.][]{Davis1985}. Therefore, the Universe's largest structures are dynamically young, having either completed their merging recently or are, in fact, still in the process of growing \citep[for example][]{Planelles2015}.

There are a number of ways to qualitatively and quantitatively characterise the large-scale environment (LSE) of galaxies and dark matter haloes (e.g. \citealt{Stoica2005, Hahn2007, Zhang2009, Sousbie2011, Hoffman2012, Cautun2013, Leclercq2016} and see \citealt{Libeskind2018} for a review of these). One class of methods is based on the construction of tensors which characterise the tidal field, specifically the Hessian of the potential ({\sc Pweb}) or the velocity shear ({\sc Vweb}). Such methods segment space into four dynamically distinct environments: knots, filaments, sheets, and voids. Following \cite{Hahn2007} and \cite{Hoffman2012}, we applied both methods to classify the LSE in \citet{Cui2018} (hereafter Paper I), which showed consistent results between the two methods. 

Pushing our previous analysis \citepalias{Cui2018} into high redshifts, we are able to present a full picture of how the LSE forms and evolves. Further, the next generation of telescopes, such as the {\it Square-Kilometre Array}\footnote{\url{https://www.skatelescope.org/}}(SKA), will be able to probe HI galaxies out to redshift $z \approx 2$ \citep{Abdalla2015}. The Hot Universe Baryon Surveyor\footnote{\url{http://heat.tsinghua.edu.cn/~hubs/en/index.html}} (HUBS) is designed to probe the warm-hot intergalactic medium (WHIM) through a soft X-ray telescope and will be able to map the warm-hot gas on large scales. In \citetalias{Cui2018}, we found that the gas web is an unbiased tracer of the dark matter web. Is this statement still valid at high redshift? And how are the baryons distributed in these LSE?

In the literature, \cite{Cautun2014} studied the evolution of the cosmic web with the {\sc NEXUS} code. Using a dark-matter-only simulation, they found that the cosmos is dominated by tenuous filaments and sheets at early times. During subsequent evolution, after the first small structures merge together, the present-day web is dominated by fewer, but much more massive, structures. Using hydro-simulations with cooling, but no star forming nor feedback schemes, \cite{Zhu2017} investigated the evolution of the cosmic web. They found that sheets appear early; filaments supersede sheets as the primary collapsing structures from $z \sim 2-3$. They further found that there is relatively less baryonic matter residing in filaments and knots compared with dark matter. In the recent work of \cite{Martizzi2018}, they applied the tidal tensor method to the Illustris-TNG simulation \citep[see][for example]{Springel2018} and found the WHIM to be the dominant baryon mass contribution in filaments and knots at redshift $z=0$, but declines faster at higher redshifts ($z > 1$). Here we are going to revisit this issue, however, using a velocity based technique (i.e. {\sc Vweb}) to define the LSE. Similarly, \cite{Nevalainen2015} found that the WHIM is usually associated with the filamentary structures($\sim$70 per cent of the gas in WHIM phase), which is identified with the Bisous model \cite{Tempel2014}.

This paper is organised as follows: we first briefly present the suite of hydro-simulations as well as the sub-grid models (\S~\ref{simulation}), and then describe the cosmological structure classification method used for this work -- the Vweb (\S~\ref{method}). In Section~\ref{results} we present our findings on the evolution of these large-scale structures as well as the baryon distributions. Finally, we summarise our conclusions and comment on the results in \S~\ref{concl}.

\section{The cosmological simulations}
\label{simulation}

The same series of three cosmological simulations presented in \citetalias{Cui2018} are used in this work. These simulations, which share the same initial conditions
% \footnote{Gas particles are treated as dark matter particles in the dark-matter-only run but with smaller mass.
with box-size $410 \Mpc$ and $2\times1024^3$ particles, use a flat $\Lambda$CDM cosmology, with cosmological parameters of $\Omega_{\rm m} = 0.24$ for the matter density parameter, $\Omega_{\rm b} = 0.0413$ for the baryon contribution, $\sigma_8=0.8$ for the power spectrum normalisation, $n_{\rm s} = 0.96$ for the primordial spectral index, and $h =0.73$ for the Hubble parameter in units of $100 \Hunit$. The masses of gas and DM particles have a ratio such that to reproduce the cosmic baryon fraction, with $m_g = 7.36 \times 10^8 \hMsun$ and $m_{DM} = 3.54 \times 10^9 \hMsun$, respectively. Gravitational forces have been computed using a Plummer-equivalent softening length which is fixed to $\epsilon_{PL} = 7.5 h^{-1}$ physical kpc from z = 0 to 2, and fixed in comoving units at higher redshift. The simulations were run with the TreePM-SPH code {\sc GADGET-3}, an improved version of the public {\sc GADGET-2} code \citep{Gadget2}. We refer to the dark-matter-only simulation as the DM run. The two hydrodynamical simulations both include radiative cooling, star formation and kinetic feedback from supernovae. In one case, we ignore feedback from AGN (which is referred to as the CSF run), while in the other we include it (which is referred to as the AGN run). Radiative cooling for gas is computed for non-zero metallicity using the cooling tables by \cite{Sutherland1993} with an evolving ultraviolet background. Star formation and supernova feedback with chemical enrichment are fully described by \cite{Tornatore2007}. The model of AGN feedback used in the AGN run is the same as that adopted by \cite{Fabjan2010}, which has optimised the parameters for galaxy cluster simulations. Interested readers are referred to \cite{Cui2012a, Cui2014a} for details on the baryon models of these simulations, and to \cite{Cui2016a, Cui2017} for details on the statistical samples of galaxy clusters from them. In this work, we only focus on four distinct redshifts, i.e. $z=[2.1, 1.0, 0.6, 0.0]$. We understand that our resolution is much lower than the one employed in the latest hydrodynamical simulations. However, this set of simulations have a much larger volume which suits for our LSE studies.

\section{Web classification method -- Vweb}
\label{method}

Following \citetalias{Cui2018}, we apply the shear tensor technique {\sc Vweb} to classify the LSE of the three simulations in this work. 
% As we have shown in \citetalias{Cui2018}, the two codes are in agreement on a certain scale (not in detailed fractions). In order to not distract the attentions of this paper, we only focus on the {\sc Vweb}. We refer to \cite{Martizzi2018} for a similar analyses, but based upon a configuration space classification of the cosmic web.

The velocity shear tensor is used to classify the cosmological space into different environments. Following \cite{Hoffman2012}, this tensor is defined as
\begin{equation}
 \Sigma_{\alpha\beta} = - \frac{1}{2H_0} \left( \frac{\partial v_\alpha}{\partial r_\beta} + \frac{\partial v_\beta}{\partial r_\alpha} \right),
\end{equation}
where, $H_0$ is the Hubble constant. The eigenvalues of $\Sigma_{\alpha\beta}$ are denoted as $\lambda_i$ ($i$ = 1, 2 and 3).

In practice, the computation of the eigenvalues for the matrix is performed on regular $256^3$ grids, corresponding to a cell size of comoving $\sim 1.6 \Mpc$. The velocity of all kinds of particles is assigned to the nearby mesh points using the triangular-shaped cloud (TSC) method. Further, a top-hat smoothing is applied to these cells. For the particular analysis of gas webs, we only use the velocities from gas particles. Then, the velocity shear tensor is calculated for each grid cell by taking derivatives from nearby cells. At last, the eigenvalues of $\Sigma_{\alpha\beta}$ for each cell are given by matrix elements calculated in the previous step. The velocity or density field estimated with different methods \citep[as discussed in][respectively]{Hahn2015, Cui2008} may lead to unacceptable levels of noise and biases. However, the large cell size and additional smoothing should be able to provide us with a robust velocity field in the real space, thereby a reliable shear tensor.

As shown in the \citetalias{Cui2018}, this cell size does not play an important role. The relative mass and volume fraction differences between the three investigated numbers of cells ($128^3, 256^3, 512^3$) are basically within 10 per cent \citep[see][for similar results]{Martizzi2018}. We also note here that the mesh size should larger than the mean size of the objects found in the simulation to avoid the interiors.

Each individual cell is then classified as either `void', `sheet', `filament', or `knot' according to the eigenvalues $\lambda_1 > \lambda_2 > \lambda_3$:
\begin{itemize}
 \item[1.] void, if $\lambda_1 < \lambda_{th}$,
 \item[2.] sheet, if $\lambda_1 \geq \lambda_{th} > \lambda_2$,
 \item[3.] filament, if $\lambda_2 \geq \lambda_{th} > \lambda_3$,
 \item[4.] knot, if $\lambda_3 \geq \lambda_{th}$,
\end{itemize}
where $\lambda_{th}$ is a free threshold parameter \citep{Hoffman2012, Libeskind2012, Libeskind2013}. Following the discussion of \cite{Carlesi2014}, we set $\lambda_{th} = 0.1$, which presents better agreement to the visualised density field. As the gas and dark matter generally share the same velocity field, the same threshold ($\lambda_{th} = 0.1$) is used for the gas component in the {\sc Vweb} code. Whether $\lambda_{th}$ depends on redshift or not is still unclear. \cite{Zhu2017} tried different redshift evolution of the $\lambda_{th}$(z) and argued that the results are in quantitatively agreement with a fixed $\lambda_{th}$. Similarly \cite{Libeskind2014} showed that the directions of the shear tensor eigen-vectors are only weakly dependent on redshift. Therefore, we adopt the same value of $\lambda_{th}$ at all our investigated redshifts for simplicity.

\section{Results}
\label{results}

\subsection{Redshift evolution of the LSE}
%%%%%%%% FIG 1 %%%%%%%%
\begin{figure*}
 \includegraphics[width=\textwidth]{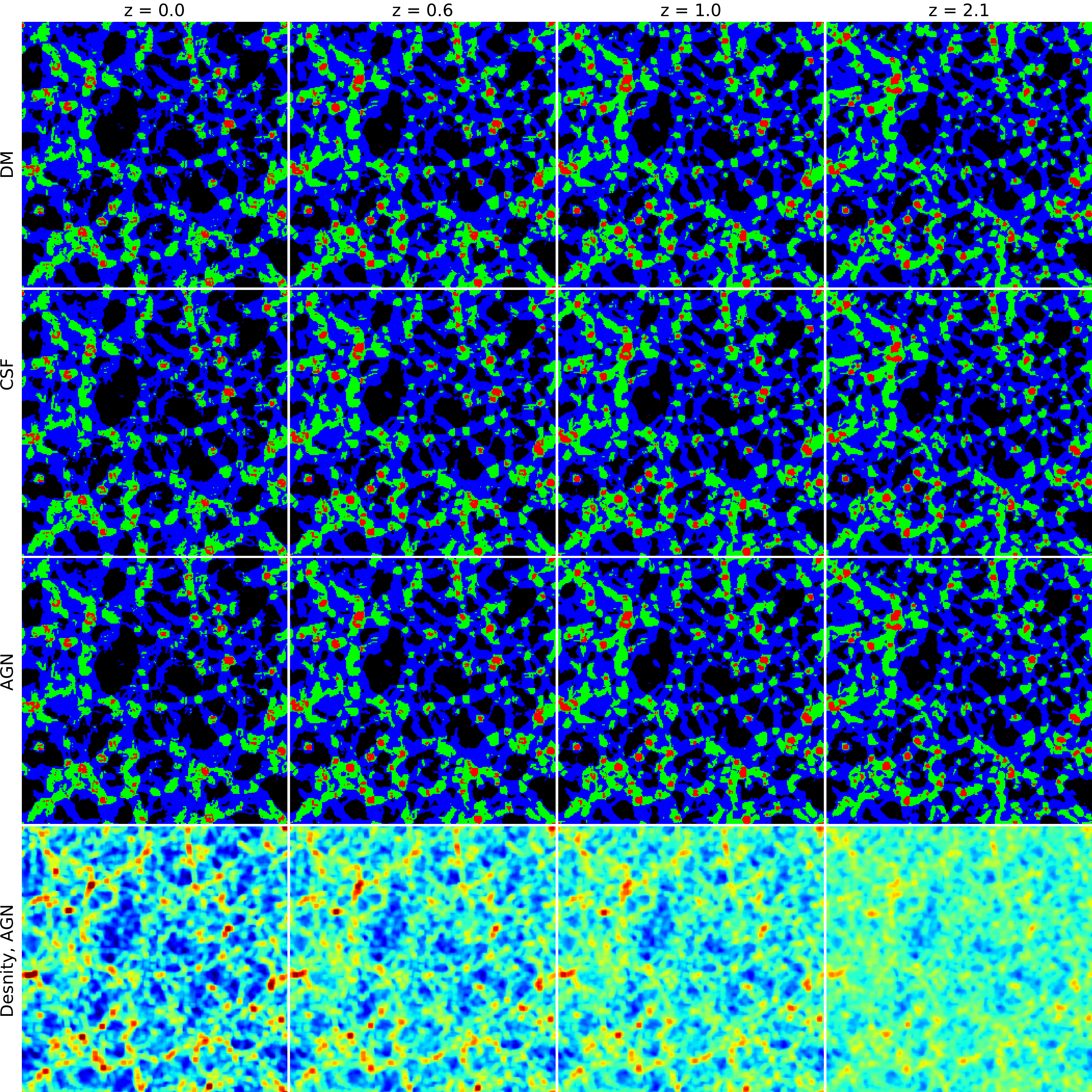}
 %\vspace*{-9mm}
 \caption{Projected large-scale structures as obtained with the {\sc Vweb} method from a slice of the whole simulation box (410$\Mpc$ along each side) of $\sim1.6 \Mpc$ thickness (one cell size): knots, filaments, sheets and void regions, which are shown in red, green, blue and black, respectively. The four columns, from left to right, show the results at $z$ = 0.0, 0.5, 1.0, 2.1, respectively. The different simulation runs, DM, CSF and AGN are shown from the first to the third row. The bottom row shows the density from the AGN run at each redshift.}
 \label{fig:show_vweb}
\end{figure*}
%%%%%%%%%%%%%%%%%%%%%%%

We first illustrate slices of projected LSE classified by the {\sc Vweb} method in Fig.~\ref{fig:show_vweb}. Each slice has a thickness of one cell size ($\sim 1.6 \Mpc$). Each cell in this projection plot corresponds to one pixel of the image. The three different runs (DM, CSF, and AGN) show an almost identical classification of the LSE at each redshift. This is not surprising as baryon processes normally leave more influences on scales smaller than the cell size used here to classify the cosmic web. While the result at $z = 0$ already are in very good agreement \citepalias[see ][for details]{Cui2018}, we expect even fewer differences between the three runs at higher redshift because the mass distribution is less clustered and hence more similar. It is clear from the density evolution (seen in the bottom row of Fig.~\ref{fig:show_vweb}) that the material is condensing into filament and knot environments. However, qualitatively these LSE show less evolution visually (cf. three upper panels). This could be due to the fact that our classification scheme is less sensitive to spatial distributions (see the volume fraction evolution in subsection \ref{sec:emv}). Similar results have been found by \cite{Martizzi2018} who used the tidal tensor method. We note here again that the same threshold ($\lambda_{th} = 0.1$ for {\sc Vweb}) is applied over all redshifts. 

\subsubsection{The evolution of the $\lambda$s}
%%%%%%%% FIG 2 %%%%%%%%
\begin{figure*}
 \includegraphics[width=\textwidth]{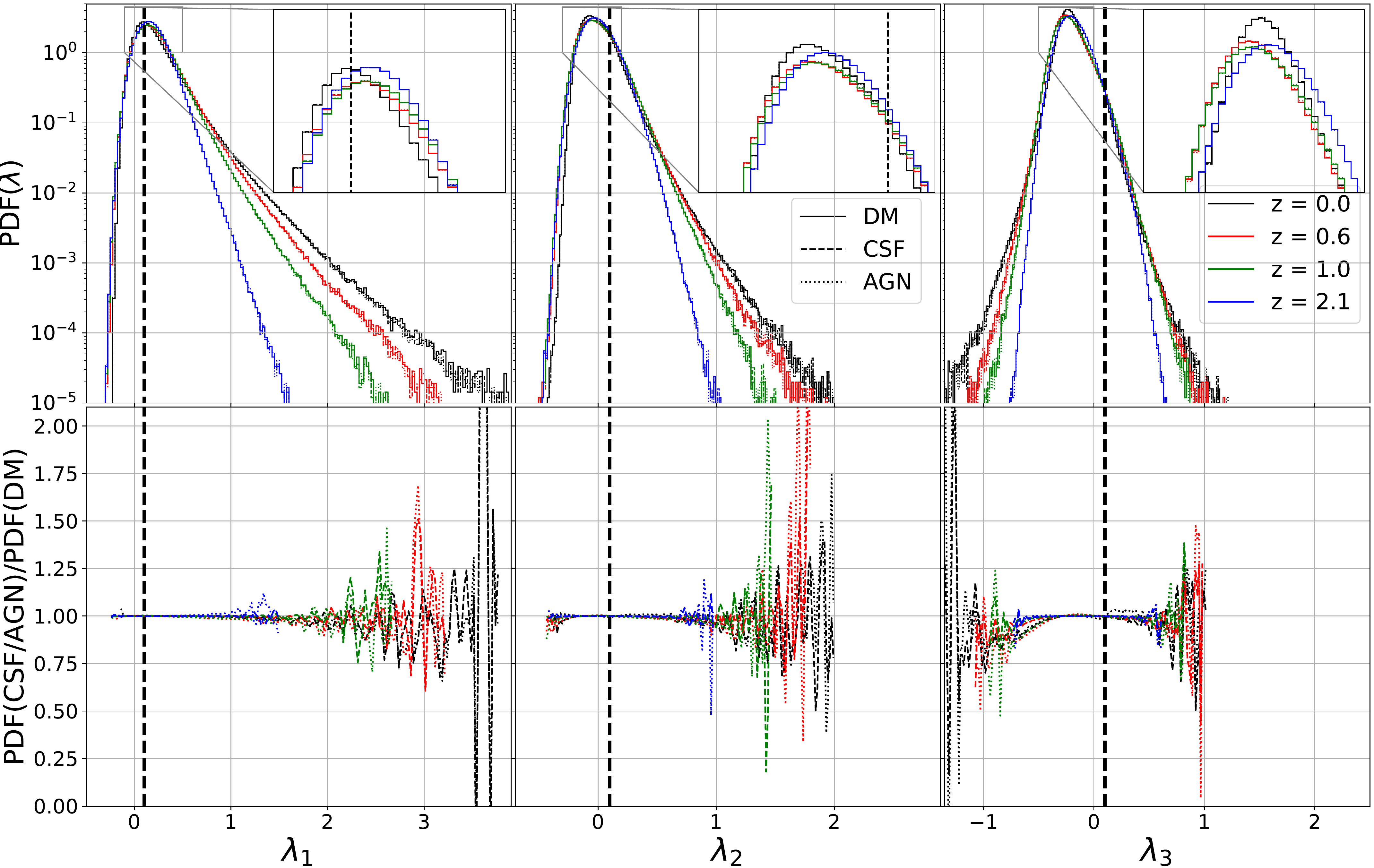}
 %\vspace*{-9mm}
 \caption{Upper row: probability distribution function (PDF) of the three $\lambda$ parameters from the {\sc Vweb} method. Lower row: the residual with respect to the DM run. As indicated in the legends inside the middle and right panels, different line style histograms indicate different simulation runs, and different colours show the results at different redshifts. The vertical dashed black line indicates the threshold $\lambda_{th} = 0.1$, which is applied to {\sc Vweb} to classify the large-scale structures. It is clear that $\lambda_1$ and $\lambda_2$ show some redshift evolution at their high value end. Weak redshift evolution for $\lambda_3$ can be seen at both maximum and minimum value ends. We highlight their peak position in the insets shown in each plot. The peak positions and the distributions around the peak show very weak redshift evolution. As indicated in the lower panel, besides the very large scatter at the tail of these PDFs, the differences between the three runs are negligible for all the redshifts and $\lambda$s.} 
 \label{fig:PDF_vweb}
\end{figure*}
%%%%%%%%%%%%%%%%%%%%%%%

In order to better quantify differences in the LSE and its redshift evolution, we present the evolution of the distributions of the three eigenvalues from the {\sc Vweb} method, which are directly used for classifying the LSE. Note, the evolution of these three $\lambda$s is directly responsible for any changes in the classification of the LSE. Actually, previous theoretical works \citep[see][for example]{Doroshkevich1989} in linear scales has predicted that the $\lambda_1$ corresponds to the formation of the "pancakes" in the high-z, the three direction contraction corresponds to the smallest eigenvalue, $\lambda_3$ with the maximum value of $\lambda_3$ determines the condensation of the clusters. Furthermore, the divergence of the velocity field is proportional to the sum of the three eigenvalues in linear regimes \citep[e.g.][]{Bernardeau1996}. The evolution of the three eigenvalues directly reflect the changes in the velocity field. The resulting probability distribution functions (PDFs) can be viewed in Fig.~\ref{fig:PDF_vweb}. The three simulations runs, which are shown by different line styles have almost the same PDFs for the three $\lambda$s at the different redshifts. In agreement with the results shown in Fig.~\ref{fig:show_vweb}, the baryon models have almost no influence on the LSE, i.e. the PDFs for DM, CSF, and AGN are practically indistinguishable (see the residuals in the lower row of Fig.~\ref{fig:PDF_vweb}). Further, the overall shape and the peak position of these PDFs for all three $\lambda$s show a very weak redshift evolution, apart from the tails, where the maximum values of $\lambda_1$ and $\lambda_2$ increase as redshift decreases. But there is almost no change for the minimum values. Because the classification of void is solely depending on $\lambda_1$ (i.e. $\lambda_1 < \lambda_{th}$ for void), the left panel clearly indicates a less significant redshift evolution from redshift $z = 2.1$ for the volume fractions of the LSE \citep[see also][]{Martizzi2018} with a negligible contribution from the baryon models.
The width of the PDF for $\lambda_3$ gets broader as redshift decreases, i.e. the maximum value increases and the minimum value decreases. This indicates that more matter is flowing into the knot environment.

\subsubsection{The evolution of the mass and volume fractions}\label{sec:emv}
%%%%%%%% FIG 3 %%%%%%%%
\begin{figure*}
 \includegraphics[width=\textwidth]{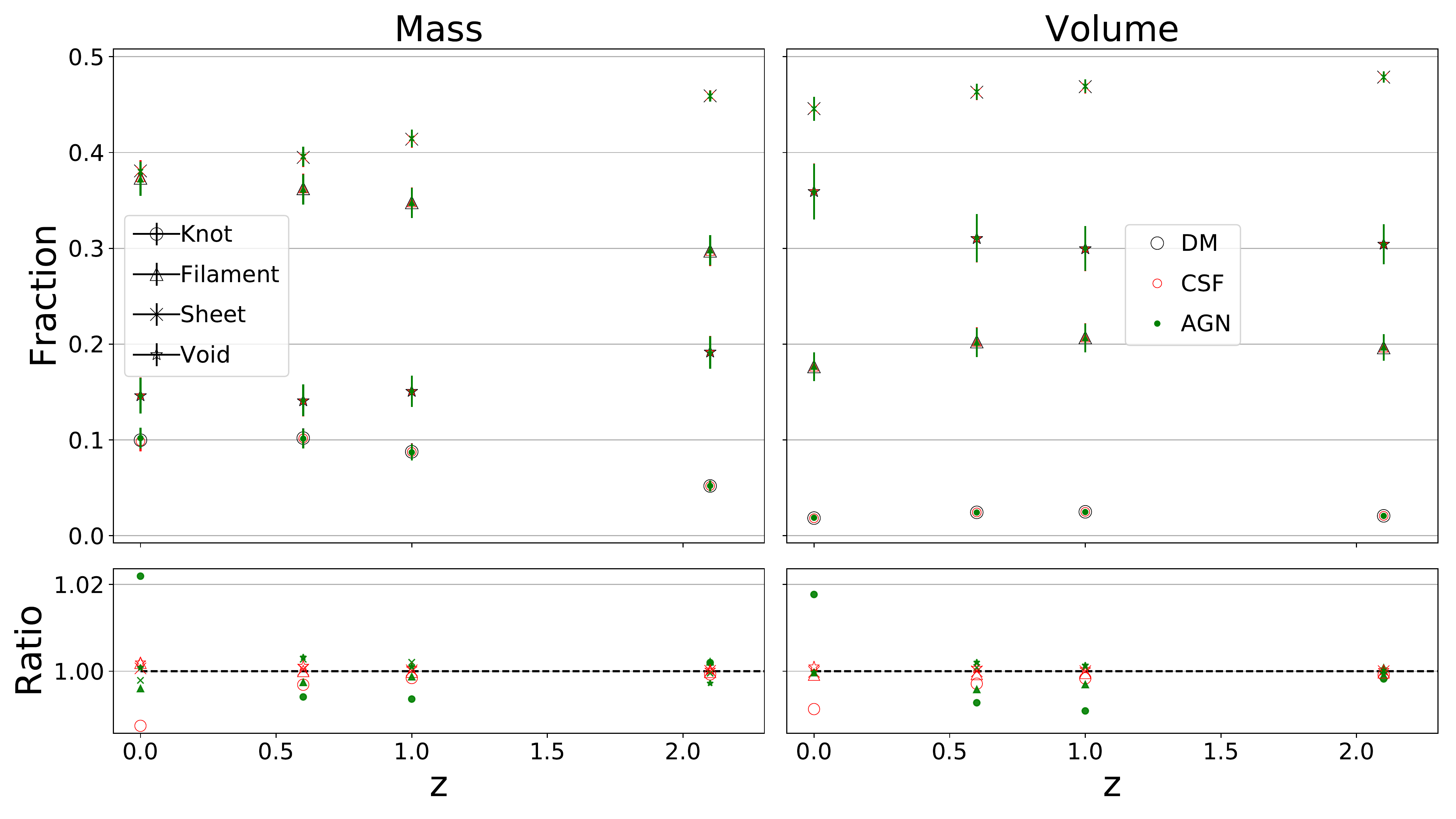}
 %\vspace*{-9mm}
 \caption{Mass and volume fractions (left and right panels respectively) of different LSE as a function of redshift from the {\sc Vweb} method. As indicated in the legend on the left-hand side panel, the four LSE (knots, filaments, sheets and voids) are shown by different type of symbols, while different simulation runs (DM, CSF and AGN) are represented by symbols with different colours and sizes. The void has an increased/decreased volume/mass fraction towards redshift $z=0$. Both filaments and knots acquire mass from sheets and voids. See the paragraph for details of estimating the error bars.
 In the lower panels, we show the relative difference between the two hydro runs and the DM run. Again, larger red symbols are for the CSF run and smaller green symbols are for the AGN run. The relative difference between the two hydro and the DM runs is very small: < 2 per cent at $z = 0$; < 1 percent at $z >= 0.6$.}
 \label{fig:FMV_vweb}
\end{figure*}
%%%%%%%%%%%%%%%%%%%%%%%

In Fig.~\ref{fig:FMV_vweb} we further explore the redshift evolution of the LSE by showing the mass (left panel) and volume fraction (right panel) of each web type (i.e. knot, filament, sheet, void) as a function of redshift. The error bars are estimated as this: the whole simulation box is equally separated into 8 small boxes with size of 205$\Mpc$; these fractions (e.g. mass fraction, volume fraction) are calculated individually in each sub-box; the error of each fraction is the standard deviation of the 8 sub-box.

For the knots environment, its volume fraction is always around 2 per cent, while its mass fraction increases towards $\sim 10$ per cent at $z = 0$. The volume fraction for the filament environment remains unchanged at the level of $\sim 20$ per cent with a slightly lower fraction at $z = 0$. On the other hand, the filament's mass fraction increases from about 30 per cent at $z=2.1$ to $\sim$40 per cent at $z=0.0$. As for the sheet environment, both its corresponding mass and volume fractions decrease with decreasing redshift. As expected from theoretical studies, the void environment occupies most of the space, but looses its mass as matter flows from low-density to high-density regions. The overall evolution of the LSE we obtain here is similar to the findings of \cite{Cautun2014, Zhu2017, Martizzi2018} who used different classification methods and different values of the threshold $\lambda_{th}$. We do not expect an exact match to their results because these fractions precisely depend on the classification method, smoothing scales, etc. \citep[see more discussions in][]{Libeskind2018}.

Even tough it is already clear from the top panels that these fractions are very close to each other when considering the three DM, CSF, and AGN runs, we further quantify the relative differences between the two hydro runs and the DM run in the lower panels by dividing the fractions for each web type and in each hydro model by the value for that web type in the DM run. At $z = 0$, both mass and volume fractions in the knot environment show the largest disagreement, around 2 per cent, which is negligible. The difference in other environments at $z = 0$ and the fraction differences in all environments at $z > 0$ are within 1 per cent. This is mainly because knot environments are the densest and are thus most susceptible to the dynamics of the baryons. 

In agreement with previous figures, the baryons leave almost no effect on both fractions over all investigated redshifts. We note here that the error bars in these fractions are generally very small, only a few percents for the largest value. These fraction differences are even smaller at higher redshifts, which can be understood as the accumulation of baryon processes and the matter distribution is identical at the initial redshift for the three simulations. Furthermore, as redshift increases, the fixed co-moving scale examined here ($\sim 1.6 \Mpc$) represents a quasi-linear length scale, especially with respect to the density field. This can be seen by examining the rms of the density field computed on this scale at $z=2$ and at $z=0$ ($\delta_{\rm rms}\approx 0.8$ and $\delta_{\rm rms}\approx 3.0$, respectively, see Fig. 4 of \citealt{Libeskind2014}).

\subsubsection{Consistency between the dark-matter and gas cosmic web}\label{sec:consistency}
%%%%%%%% FIG 4 %%%%%%%%
\begin{figure*}
 \includegraphics[width=\textwidth]{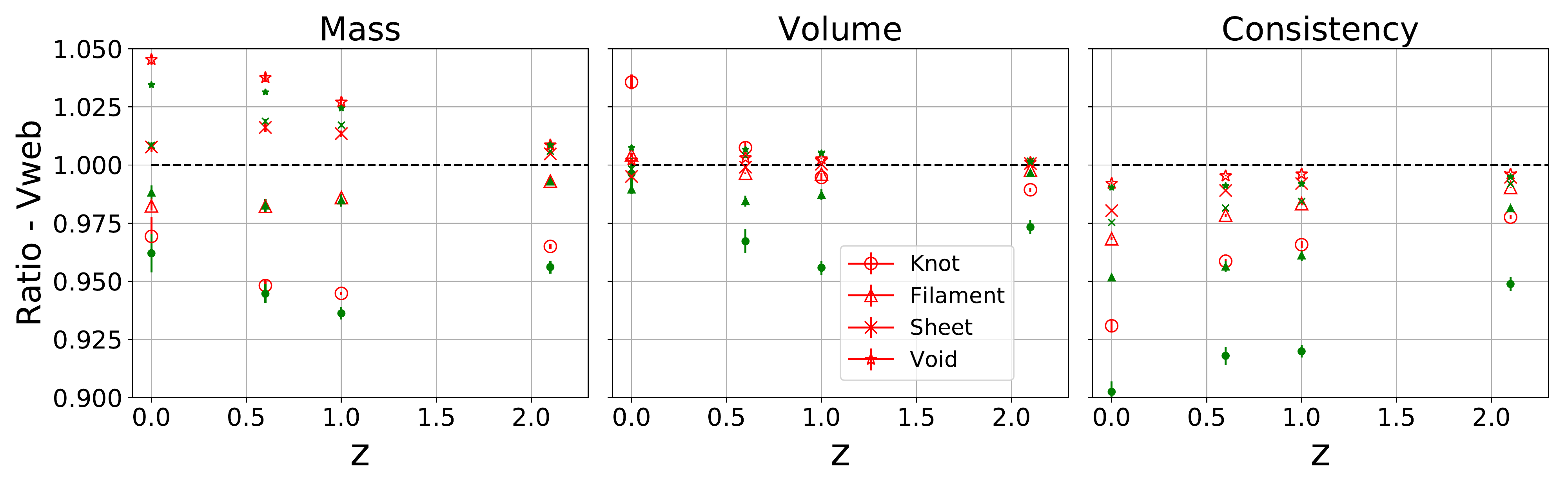}
 %\vspace*{-9mm}
 \caption{Ratio of the mass (left panel), volume (middle panel) and consistency (right panel) fractions as a function of redshift between the results that use all matter and the results that use only the gas component to classify the LSE. As indicated in the legend on the middle panel, different symbol styles are for the four LSE, with larger red open symbols for the CSF run and smaller green filled symbols for the AGN run. See Section\ref{sec:emv} for details of estimating the error bars.}
 \label{fig:FMV_diff}
\end{figure*}
%%%%%%%%%%%%%%%%%%%%%%%

One important finding of \citetalias{Cui2018} is that gas filaments are an unbiased tracer of the large-scale structures -- at least at redshift $z=0$ (see the right column of Fig.~5 in there). Therefore, it is interesting to ask whether that conclusion is still valid at high redshift or not. We now extend our previous analysis to higher redshifts and show in Fig.~\ref{fig:FMV_diff} the relative difference for the mass (left-hand side panel), volume (middle panel) and consistency (right-hand side panel) fractions between the results from the gas-only component and the results from total matter. The consistency fraction is defined as the number of cells that are identified as the same web type in both the gas-only and total matter results divided by the the total number of cells in that web environment found by considering the total matter result. This consistency level best quantifies how the two results agree with each other.

In the left panel, the ratio (the mass fraction classified by the gas component divided by the mass fraction computed from the total matter) for void and sheet environments is always greater than unity (from $\sim$ 1 per cent to 5 per cent, depending on the redshift). 
This could be because the gas is more smoothly distributed at semi-linear scale (such as the cell size used in this paper) than the collision-less dark matter. Gas follows dark matter to drop into its potential. However, it has additional supports from the pressure to slow down its clumpiness. Therefore, if only the gas component is used to classify the LSE instead of the total, which is dominated by dark matter, more gas tends to be assigned into void and sheet environments, leaving a lower amount allocated in the knot and filament environment. Consistently, the volume fraction is also tiny higher in the void and sheet environments. Nevertheless, it is worth to note that these fractions are very small, which is almost negligible.
The agreement (within 1 per cent) for the mass fractions in filaments, sheets and voids becomes better at high redshift, but not for the mass fraction in knots where the relative difference is at about 3.5 - 5 per cent. For all ratios, the error bars are very small which indicates these results are robust.
% This could be caused by the smooth distribution of gas, which has additional pressure support besides tracing the gravitational potential of dark matter. Therefore, compared to dark matter, more gas reminds in the void and sheet environments. On the contrary, less mass is allocated into the knot and filament environments when only the gas component is used to classify the LSE. The agreement for the mass fractions in filaments, sheets and voids becomes better at high redshift, but not for the mass fraction in knots.

Unlike the ratio for the mass fraction, the volume ratios in the middle panel show no clear redshift evolution. Besides the volume fraction ratio in the knot environment (from the CSF run at z = 0 and from the AGN run at higher redshift), the difference is much smaller than the mass fraction differences, within about 2 per cent. 

The consistency fraction in the right panel is always smaller than 1, which means that the same LSE are identical with both classifications from gas-only component and total matter. It is clear that there is a better agreement at higher redshift for all LSE. Again, the knot environment shows the largest disagreement compared with the other environments. This is because baryon physics plays a more important role in denser environments and on smaller scales. This is especially true for the AGN run, because the gas distribution in knot environments is largely affected by AGN feedback. 

Interestingly, for the filament environment, both mass and volume difference fractions are very small (within about $\sim 2$ per cent), which is similar to the $z$=0 results. Over 95 per cent of the cells are cross identified in the gas-only filaments and total matter filaments. These fractions and the consistency ratio show better agreement at higher redshift. Therefore, we can conclude that our previous finding, i.e. that the gas component is an unbiased tracer of the large-scale structure, as expected, is still valid for all redshifts -- with an even better agreement at the highest redshift $z$=2.1. 

\subsection{The evolution and distribution of baryons}
The baryons, which occupy less than 5 per cent of the total energy content in the Universe, are the only component of the Universe which can be directly observed. The gas component, which occupies about 90 per cent of total baryons \citep{Nicastro2018}, can be roughly separated into: 1) cold gas (T < 10$^5$ K) in diffused or condensed environments; 2) warm-hot gas (10$^5$ < T < 10$^7$) in the intergalactic medium (WHIM); 3) hot gas (T > 10$^7$) mostly in halos \citep[e.g.][]{Dave2001,Cen2006,Tornatore2010,Planelles2018}. It is believed that the WHIM, which is very hard to be detected, plays a central role to solve the missing baryon problem, namely the mismatch (about one-tenth) in the apparent baryon content between observations at low and high redshift \citep[][etc.]{Persic1992,Fukugita1998,Cen1999,Bregman2007,Shull2012, Nicastro2018,Fang2018}. Recent works by \cite{deGraaff2017,Tanimura2019} have applied different methods to search for the missing baryons and found that the WHIM tends to be located in the filaments and between pairs of merging systems.
Therefore, in this Section, we detail the evolution and the distribution of baryons and hope to shed some light on this issue from the theoretical point of view.

\begin{table*}
    \centering
    \caption{The cosmological gas and stellar mass fractions in percentiles with respect to the total baryon mass. The fractions from the AGN run are shown in the first followed by the results in brackets from the CSF run. The results from \citet{Nicastro2018} are at z < 0.5. We further note here that the WHIM fraction from \citet{Nicastro2018} has a slightly different upper temperature limit (10$^{6.2}$ versus 10$^7$). The fractions from \citet{Haider2016} are coming from the Illustris simulation \citep{Vogelsberger2014}. The same method introduced in Section\ref{sec:emv} is used to estimate these error bars.}
    \begin{threeparttable}
        \begin{tabular}{|c|c|c|c|c|}
        \hline
            Redshift & hot gas & WHIM & cold gas & star \\
            \hline
            \cite{Nicastro2018}\tnote{a} & 9$\pm$4.5 & $\gtrsim$24 \& $\lesssim$55 & 29.7$\pm$11 & 7$\pm$2\\
            \cite{Haider2016}\tnote{b} & 6.5 & 53.9 & 32.8 & - \\
            z = 0   & 4.6$\pm$0.7 (4.6$\pm$0.1) & 41.3$\pm$1.1 (38.3$\pm$1.0) & 50.9$\pm$0.1 (50.6$\pm$0.2) & 3.2$\pm$0.1 (6.5$\pm$0.2)\\
            z = 0.6 & 2.4$\pm$0.4 (1.1$\pm$0.2) & 34.9$\pm$1.1 (29.7$\pm$1.1) & 60.2$\pm$0.1 (65.0$\pm$0.1) & 2.5$\pm$0.1 (4.2$\pm$0.2)\\
            z = 1.0 & 1.3$\pm$0.2 (0.3$\pm$0.0) & 28.7$\pm$1.1 (21.9$\pm$1.0) & 68.1$\pm$0.1 (74.8$\pm$0.1) & 1.9$\pm$0.1 (2.8$\pm$0.1)\\
            z = 2.1 & 0.2$\pm$0.0 (0.0$\pm$0.0) & 10.8$\pm$0.5 (6.2$\pm$0.4) & 88.2$\pm$0.1 (92.9$\pm$0.0) & 0.8$\pm$0.0 (0.8$\pm$0.0)\\
            \hline
        \end{tabular}
        \begin{tablenotes}
        \item[a] These fractions are estimated at $z<0.5$.
        \item[b] These fractions are with respect to the total gas mass at $z=0$.
        \end{tablenotes}
    \end{threeparttable}
    \label{tab:cos_bfrac}
\end{table*}

We first investigate the cosmological baryon fractions in our two hydrodynamical simulations and compare them with the observational results from \cite{Nicastro2018}, which are presented in Table~\ref{tab:cos_bfrac}. Our simulations give a similar WHIM mass fraction as the observations, but the cold gas fractions are a bit higher. The effect of the AGN feedback does not only reduce the total stellar mass by almost half of the CSF run \citep[e.g.][]{Planelles2013b}, but it also heats a certain amount of cold gas into the warm/hot phase. The difference of the warm/hot gas mass fraction between the our two simulations is much larger at the two middle redshifts. Therefore, it results in a larger difference in the cold gas fraction as well. This indicates that the AGN feedback plays a more important role (more powerful) at higher redshifts than at z = 0, when most of the AGN feedback is seized due to the galaxies being quenched in the simulation.

It is worth noting that the stellar mass fraction formed in the CSF run is comparable to \cite{Nicastro2018}. However, the simulation with AGN feedback, which is believed to be necessary to reproduce the correct stellar mass function, has much lower stellar mass fraction than that seen in observation. This contradiction found in the AGN run, hence such theoretical models (and their numerical implementation), implies that the subgrid baryon models are still challenging. 
%We note here that different cosmological parameters are used as compared to \cite{Nicastro2018}. where they are based upon \cite{Planck2016}. The slightly larger $\Omega_b = 0.0487$ in \cite{Nicastro2018} (ours is $\Omega_b = 0.0413$) may be responsible for the higher stellar mass fraction: a higher baryonic fraction will increase the gas density and eventually lead to more star formation.

Compared to the results from \cite{Haider2016}, their WHIM fraction is about 12 per cent higher than ours with a much lower mass fraction ($\sim$18 per cent) in cold gas. Note that their gas mass fractions are normalised to the total gas mass instead of total baryonic mass. Further, it is known that the AGN feedback in Illustris is even powerful that gives much more hot gas. Similarly, \cite{Martizzi2018} also report that the mass in the WHIM in Illustris is $\sim$23\% larger than in IllustrisTNG. 

\subsubsection{The evolution of the gas density}

%%%%%%%% FIG 5 %%%%%%%%
\begin{figure*}
 \includegraphics[width=\textwidth]{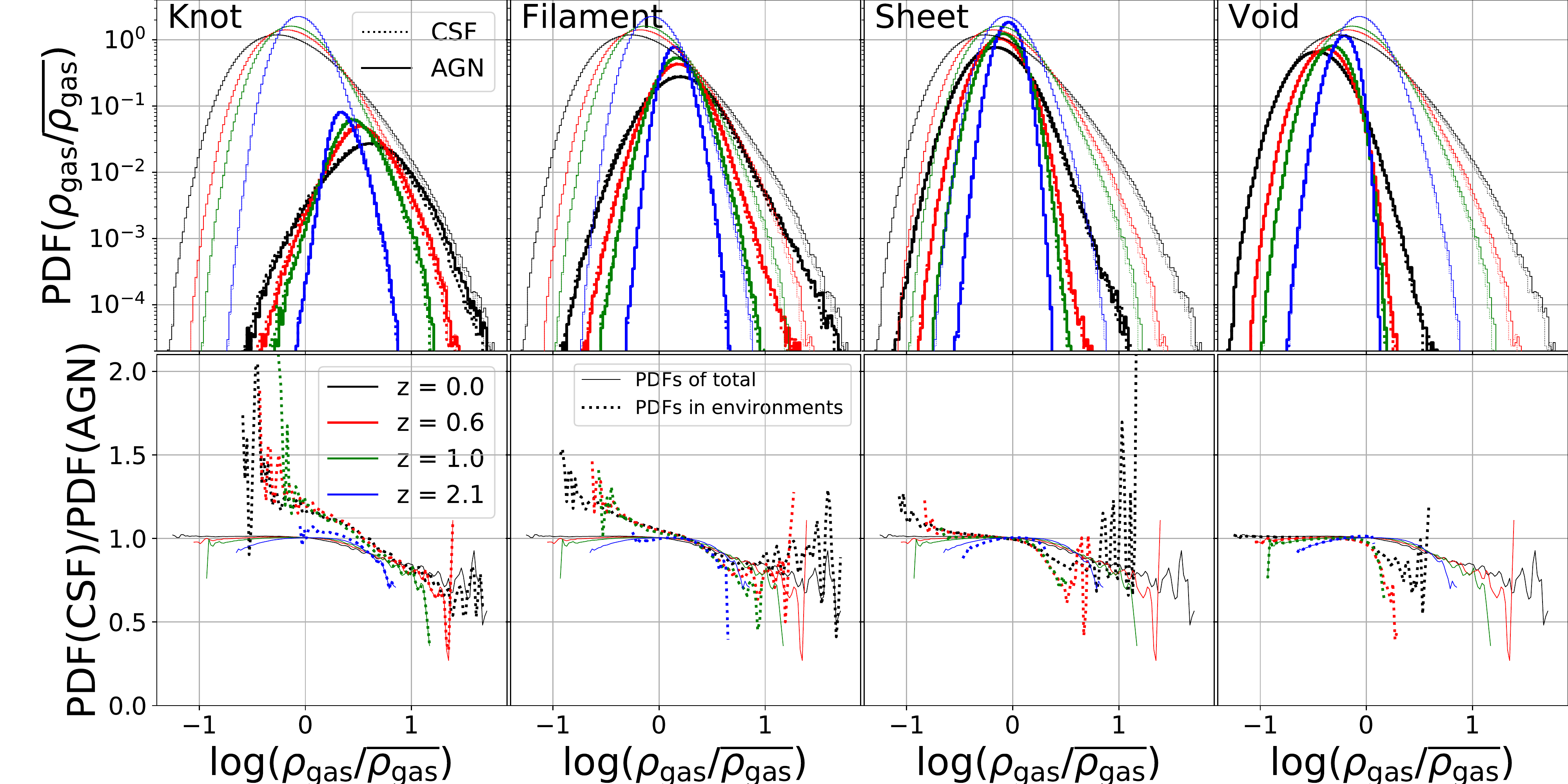}
 %\vspace*{-9mm}
 \caption{Upper row: PDFs of the gas density $\rho$ in each cell at different redshifts and LSE; Lower row: the relative difference between the two runs. From left to right panels, we show the PDFs of the gas density (normalised by the mean) in knot, filament, sheet and void environments, respectively. Note that the thin lines are for the total gas density, which are identical for the same redshift in different panels. Different line colours represent different redshifts as shown in the legend in the top-right panel. The solid line shows the result from the AGN run while the dotted line is from the CSF run. The thick lines show the density PDFs (re-normalised by their numbers to match the PDF from the total gas PDFs) inside the corresponding LSE. The relative differences between the two runs are shown in the lower row by thin solid lines for the results from total gas and thick dotted lines for the results in each environment. This figure highlights the evolution and distribution of the gas density in the whole simulation as well as in the different LSE.}\label{fig:pdf_gas}
\end{figure*}
%%%%%%%%%%%%%%%%%%%%%%%

We now examine the evolution and distribution of the gas in the simulation as well as in different LSE by showing its density PDF in Fig.\ref{fig:pdf_gas}. Instead of directly using the density of each gas particle in the simulation, we recalculate it inside each cell.\footnote{We use the same grid for the density calculation as already used for the cosmic web classification with {\sc Vweb}.} As it is well known, the distribution of density in the Universe is expected to be close to log-normal \citep[see][and references therein]{Coles1991}. The PDF of the log of normalized gas density therefore has its peak around 0 with a narrow width at $z = 2.1$. As redshift decreases down to zero, the PDF profile gets broader with the peak shifting to smaller values. This can be easily understood as the gas moves from low density regions (voids) to higher dense regions (knots), which will make the void regions even emptier (decreased min density value) and the knot region more compact (increased max density value). This behaviour is well known at least since the seminal work of \cite{Coles1991} \citep[although see][for a modern view of the distribution of cosmic density]{Klypin2018}, yet it is worth noting that the two hydrodynamical simulations share the same PDF distributions for low gas density. There is a very small discrepancy (slightly higher density in the AGN run than in the CSF run) at the higher density tail especially at low redshift. That could be caused by the AGN feedback stopping the star formation, leaving more gas inside the halos.

It is not surprising to see the gas with highest density tends to be located in the knot environment. From the filament to the sheet and the void environments, the PDFs of the gas density gradually shift from higher density to lower density with the most under-dense gas located within the void region. Similar to the PDFs of total gas density, the curve of these PDFs in different LSE also have the shape of a log normal distribution with smaller widths at high redshift and broader widths at low redshift. The redshift evolution of these PDFs in different LSE is generally following the total PDFs at the corresponding density range.

Seeing from the lower panel, the AGN feedback seems to leave no influences on the low density gas from the total. The CSF run tends to have less high dense gas compared to the AGN run. This can be understood as the AGN feedback stops the star formation and has more gas left without condensed to stars. It is not surprising to see that the gas PDF in each environment basically follows the total PDFs. It is worth to note that the CSF run tends to have relatively more low-dense gas in the knot and filament environments than the AGN run. This is consistent with the previous explanation as the knot and filament environments can be strongly affected by the AGN feedback than sheet and void environments.

\subsubsection{The evolution of the gas temperature}
%%%%%%%% FIG 6 %%%%%%%%
\begin{figure*}
 \includegraphics[width=\textwidth]{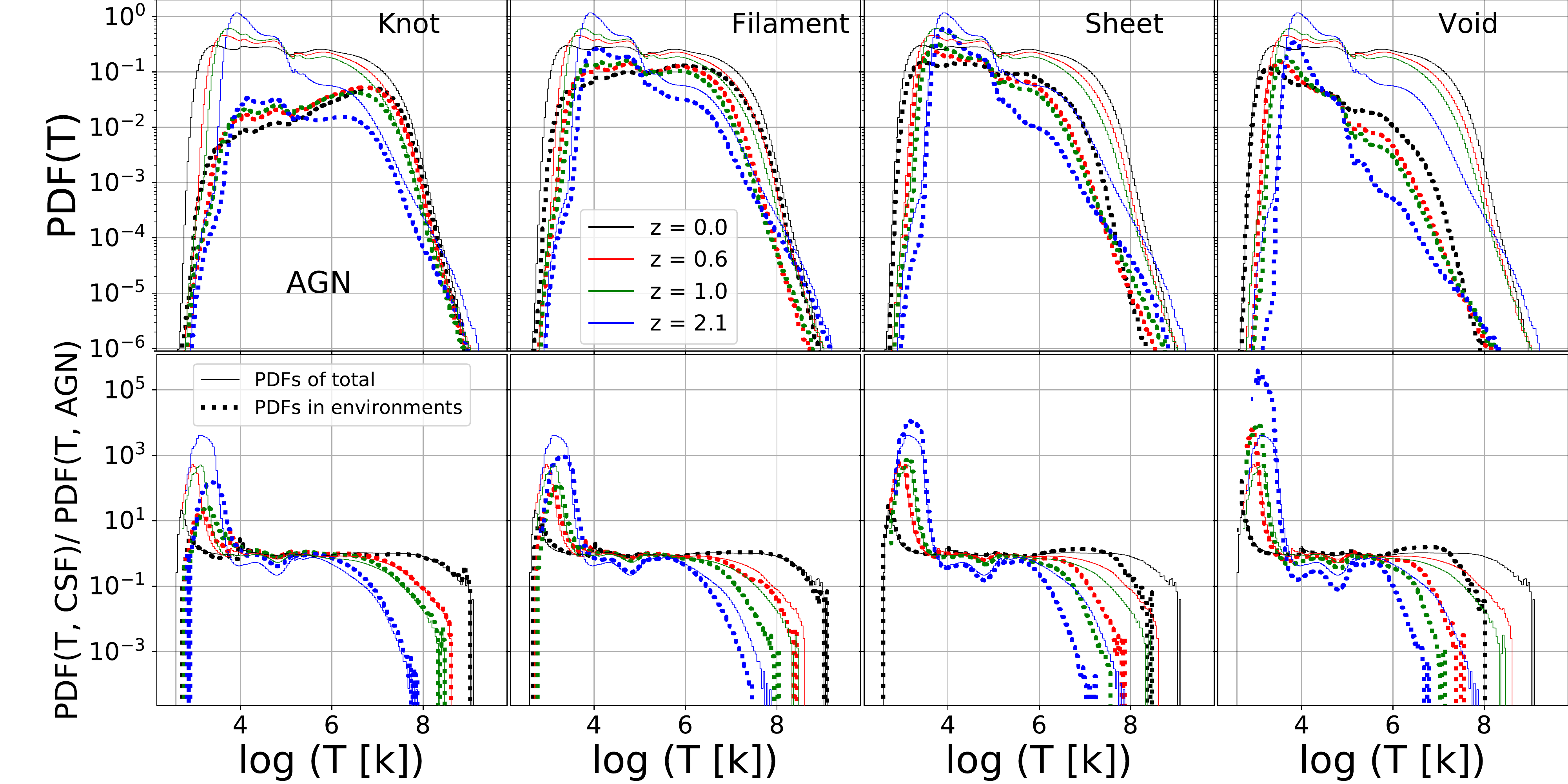}
 %\vspace*{-9mm}
 \caption{Similar to Fig.~\ref{fig:pdf_gas}, the PDFs of the gas temperature in each cell at different redshifts and in different LSE from only the AGN run in upper row with the relative difference in lower row. As indicated in the top-right corners on the panels of top row, the results in the four environments (i.e. knot, filament, sheet and void) are shown from left to right, respectively. Note that the thin lines are for the total gas temperature, which are identical for the same redshift in different panels. Different line colours represent different redshifts as shown in the legend. The dotted thick lines show the density PDFs (re-normalised by their numbers to match the PDF from the total gas PDFs) inside the corresponding LSE. Instead of showing the results from the CSF run, we present the relative difference between the two runs in the lower row. Again, the thin solid lines show the relative difference from the total while the thick dotted lines show the ratio in each environment. This figure highlights the evolution and distribution of the gas temperature in the whole simulation as well as in the different LSE.}\label{fig:pdf_gas_temp}
\end{figure*}
%%%%%%%%%%%%%%%%%%%%%%%

The distribution and evolution of the gas temperature, which is calculated as the mass-weighted mean in the cells, are presented in Fig.~\ref{fig:pdf_gas_temp}. We only show the PDFs from the AGN run in upper row with the relative difference in the lower row. Similar to the total gas density PDFs, the total gas temperature distribution evolves along the redshift with a decrease of the peak ($\sim 10^4 $[k] at $z$=2,1) and an increase of both high and low temperature gas. In knot environment, we find a relatively larger spread of temperature than for the density due to the co-existence of the cold and hot gas in these regions. The gas temperature distribution in the filament environment is also rather wide. However, the low temperature gas starts to contribute more. In both sheet and void regions, the cold gas is the dominant component. There are very few cells with higher temperature $> 10^6$ K, especially at higher redshift.

Unlike the PDF for the gas density in Fig.~\ref{fig:pdf_gas}, there are clear differences between the CSF and the AGN run. The AGN run tends to systemically have higher temperature gas (about a factor of 2 for $T > 10^4$ with redshift dependence) than the CSF run. While the CSF run exceeds the AGN run by orders of magnitude for the gas at temperature of $\sim 1000$ K. Not surprisingly, the same level of disagreement between the two runs is also showing in each of our LSE. This indicates that the AGN feedback is more efficient in changing the gas temperature than the actual gas distribution.

\subsubsection{The evolution of the gas phase diagram}
%%%%%%%% FIG 7 %%%%%%%%
\begin{figure*}
 \includegraphics[width=\textwidth]{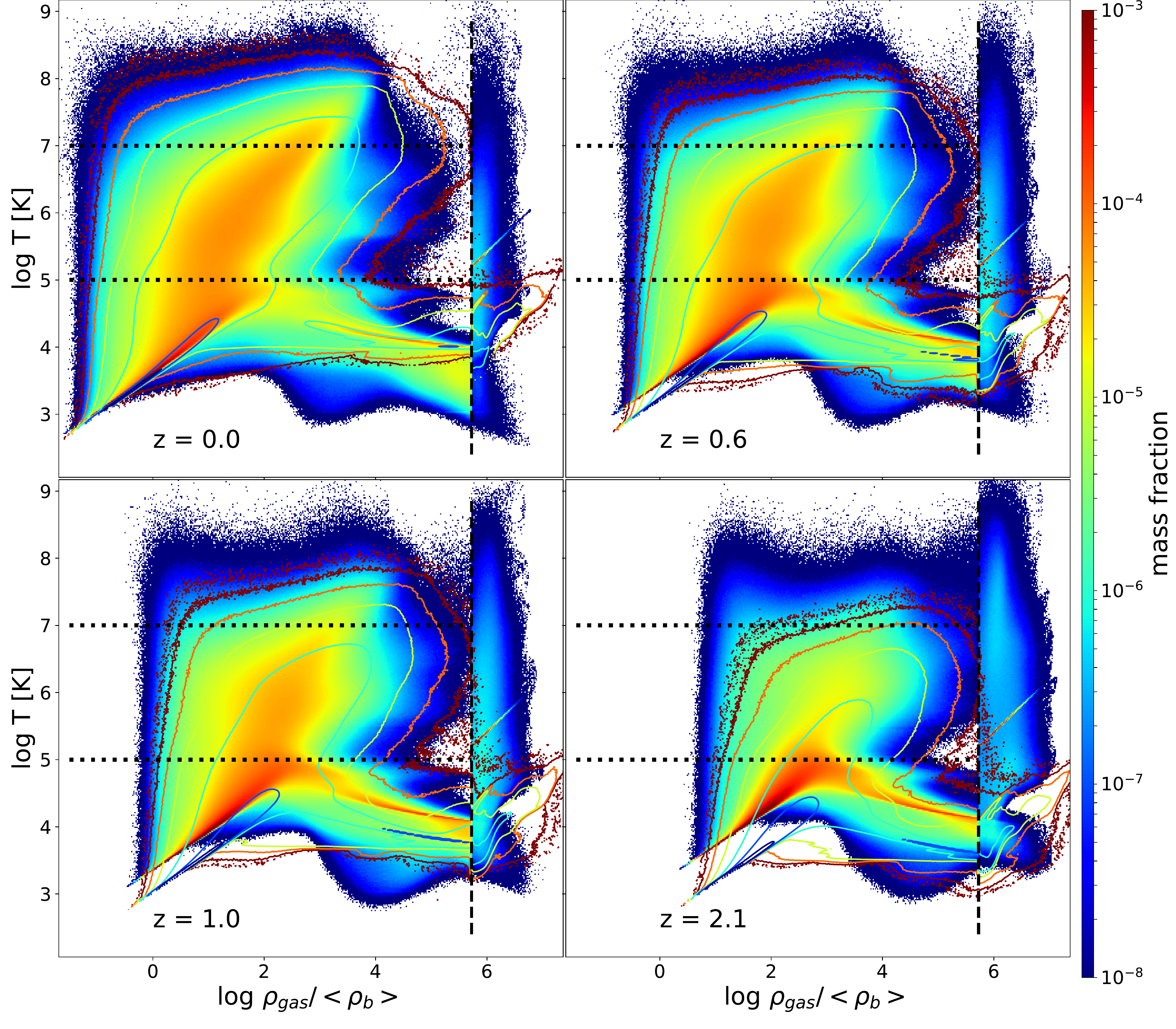}
 %\vspace*{-9mm}
 \caption{Gas density -- temperature diagram at the four different redshifts from top-left to bottom-right panels. The results from the AGN run are shown by the colour-map with the colour-bar indicating the mass fraction on the right-hand side. The results from the CSF run are presented with the contours lines in reversed colour. The vertical and horizontal lines are for the star formation threshold and gas temperature separations.}
 \label{fig:gas_rt}
\end{figure*}
%%%%%%%%%%%%%%%%%%%%%%%

We present the gas phase diagram at different redshifts in the four panels of Fig.~\ref{fig:gas_rt}. The gas density -- temperature distribution is separated into 4 different regions by the black vertical and horizontal lines. The vertical lines indicate the limitation of gas density for star formation. Star forming gas particles are assigned to the cold gas phase ignoring their temperature. The larger distribution of the temperature for the star forming gas is because the gas is treated as multi-phase, so as to provide a sub-resolution description of the interstellar medium. For our considered regions below the gas density limitation, the gas is simply separated by its temperature into three parts: hot gas, WHIM and cold gas. It is clear that the cold gas dominates the mass fraction at $z = 2.1$. As redshift decreases, more and more gas is heated up into the WHIM and hot region. Detailed fractions can be found in Table~\ref{tab:cos_bfrac}. There are noticeable differences between the two runs, especially at high redshifts and in the low temperature region. This could be due to the fact that cooling in the CSF runs is more efficient without the feedback from AGN. It is interesting to see that the contours are in good agreement with the colour-map at $z = 0$. This means that the AGN feedback is again more powerful at higher redshift and becomes negligible after the galaxies are quenched.

\subsubsection{The distribution of baryons in different environments}
%%%%%%%% FIG 8 %%%%%%%%
\begin{figure*}
 \includegraphics[width=\textwidth]{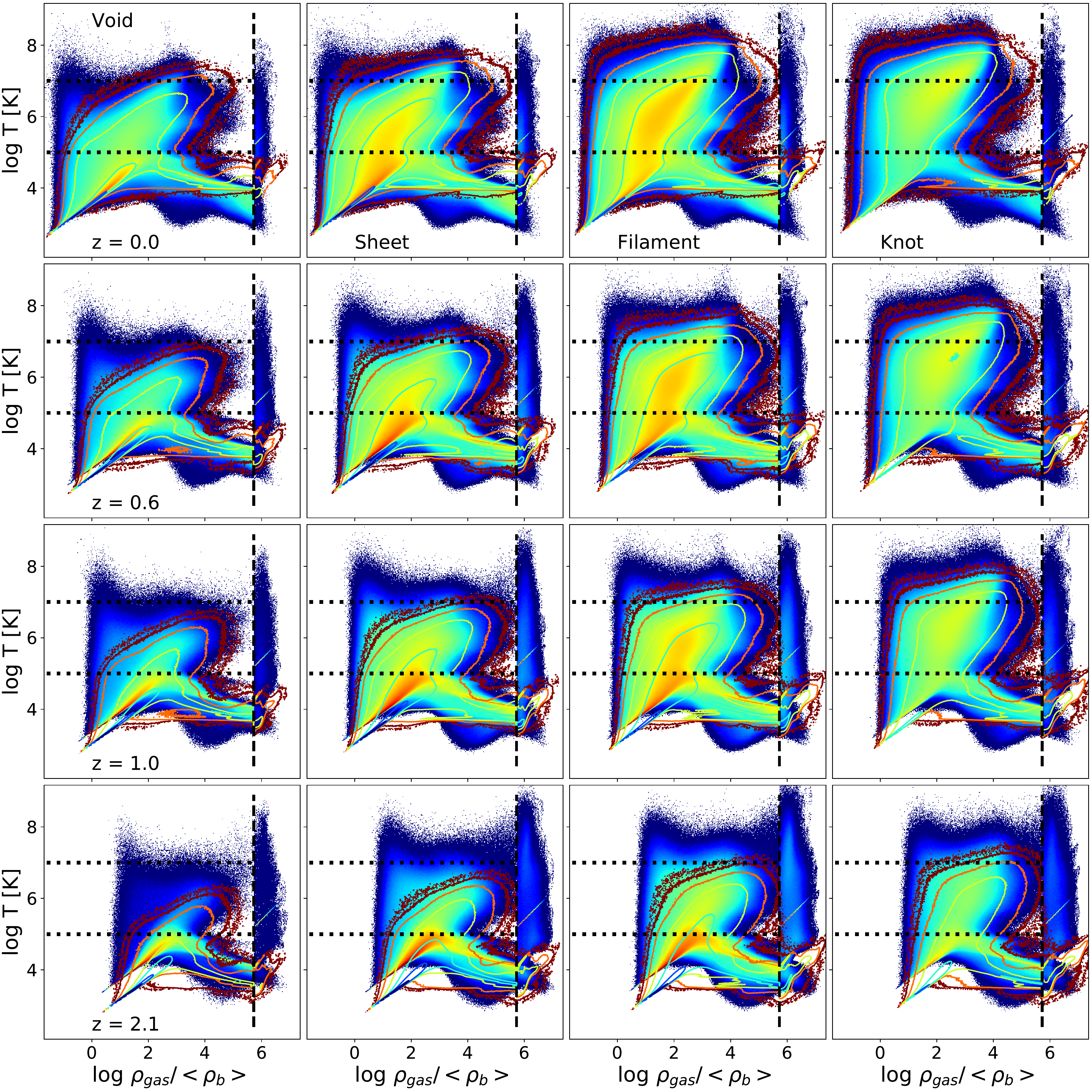}
 %\vspace*{-9mm}
 \caption{Similar to Fig.~\ref{fig:gas_rt}, the gas phase diagram in different LSE shown in the same columns. The total gas is separated into different LSE according to its position. From top to bottom panels, we show the result in the four redshifts. Again, the colour-maps are coming from the AGN run, while the contours are for the CSF run. The colour indicates the gas mass fraction in each region. Note that the contours use the reversed colour to highlight the difference.}
 \label{fig:gas_rte}
\end{figure*}
%%%%%%%%%%%%%%%%%%%%%%%

\begin{table*}
\fontsize{9}{9}\selectfont
\caption{The distribution of different baryon components in the LSE at the four redshifts. The AGN results are shown by the first value with the CSF results follow inside the brackets. These fractions are normalised to the total mass of each baryon component, i.e. the sum of each line equals 1. See Section\ref{sec:emv} for details of how these error bars are calculated.}
\label{tab:env_bfrac}
  \begin{tabular}{c|c|c|c|c}
       & Voids & Sheets & Filaments & Knots \\
       \hline
       & & z = 0 & & \\
       \hline
    $f_{\rm total\ gas}$ & 0.15$\pm$0.02 (0.16$\pm$0.02) & 0.38$\pm$0.01 (0.39$\pm$0.01) & 0.37$\pm$0.02 (0.36$\pm$0.02) & 0.10$\pm$0.01 (0.09$\pm$0.01) \\
    $f_{\rm star}$       & 0.08$\pm$0.01 (0.06$\pm$0.01) & 0.33$\pm$0.02 (0.29$\pm$0.02) & 0.44$\pm$0.01 (0.48$\pm$0.01) & 0.15$\pm$0.02 (0.17$\pm$0.02) \\
    $f_{\rm hot\ gas}$   & 0.00$\pm$0.00 (0.00$\pm$0.00) & 0.04$\pm$0.01 (0.05$\pm$0.01) & 0.46$\pm$0.03 (0.50$\pm$0.03) & 0.49$\pm$0.03 (0.45$\pm$0.03) \\
    $f_{\rm WHIM}$       & 0.05$\pm$0.01 (0.06$\pm$0.01) & 0.30$\pm$0.02 (0.31$\pm$0.02) & 0.51$\pm$0.01 (0.50$\pm$0.01) & 0.14$\pm$0.01 (0.13$\pm$0.01) \\
    $f_{\rm cold\ gas}$  & 0.25$\pm$0.02 (0.25$\pm$0.02) & 0.48$\pm$0.01 (0.48$\pm$0.01) & 0.24$\pm$0.02 (0.25$\pm$0.02) & 0.03$\pm$0.01 (0.03$\pm$0.01) \\
    \hline
       & & z = 0.6 & & \\
       \hline
    $f_{\rm total\ gas}$ & 0.14$\pm$0.02 (0.14$\pm$0.02) & 0.39$\pm$0.01 (0.39$\pm$0.01) & 0.37$\pm$0.02 (0.36$\pm$0.02) & 0.11$\pm$0.01 (0.11$\pm$0.01) \\
    $f_{\rm star}$       & 0.05$\pm$0.01 (0.03$\pm$0.00) & 0.28$\pm$0.02 (0.22$\pm$0.02) & 0.47$\pm$0.01 (0.48$\pm$0.01) & 0.21$\pm$0.02 (0.27$\pm$0.02) \\
    $f_{\rm hot\ gas}$   & 0.00$\pm$0.00 (0.00$\pm$0.00) & 0.01$\pm$0.00 (0.00$\pm$0.00) & 0.23$\pm$0.01 (0.15$\pm$0.02) & 0.76$\pm$0.02 (0.85$\pm$0.02) \\
    $f_{\rm WHIM}$       & 0.03$\pm$0.00 (0.03$\pm$0.00) & 0.25$\pm$0.02 (0.23$\pm$0.02) & 0.53$\pm$0.01 (0.52$\pm$0.01) & 0.20$\pm$0.01 (0.22$\pm$0.01) \\ 
    $f_{\rm cold\ gas}$  & 0.21$\pm$0.02 (0.19$\pm$0.02) & 0.48$\pm$0.01 (0.47$\pm$0.01) & 0.28$\pm$0.02 (0.29$\pm$0.02) & 0.04$\pm$0.01 (0.04$\pm$0.01) \\
    \hline
      & & z = 1.0 & & \\
      \hline
    $f_{\rm total\ gas}$ & 0.14$\pm$0.02 (0.14$\pm$0.02) & 0.40$\pm$0.01 (0.40$\pm$0.01) & 0.36$\pm$0.02 (0.35$\pm$0.02) & 0.10$\pm$0.01 (0.10$\pm$0.01) \\
    $f_{\rm star}$       & 0.03$\pm$0.01 (0.02$\pm$0.00) & 0.26$\pm$0.02 (0.20$\pm$0.02) & 0.48$\pm$0.01 (0.48$\pm$0.01) & 0.22$\pm$0.02 (0.29$\pm$0.02) \\
    $f_{\rm hot\ gas}$   & 0.00$\pm$0.00 (0.00$\pm$0.00) & 0.02$\pm$0.01 (0.00$\pm$0.00) & 0.22$\pm$0.02 (0.08$\pm$0.02) & 0.76$\pm$0.02 (0.91$\pm$0.02)\\
    $f_{\rm WHIM}$       & 0.02$\pm$0.00 (0.02$\pm$0.00) & 0.24$\pm$0.01 (0.21$\pm$0.01) & 0.53$\pm$0.01 (0.52$\pm$0.01) & 0.21$\pm$0.01 (0.25$\pm$0.01) \\
    $f_{\rm cold\ gas}$  & 0.20$\pm$0.02 (0.18$\pm$0.02) & 0.48$\pm$0.01 (0.46$\pm$0.01) & 0.28$\pm$0.02 (0.31$\pm$0.02) & 0.04$\pm$0.01 (0.05$\pm$0.01) \\
    \hline
      & & z = 2.1 & & \\
      \hline
    $f_{\rm total\ gas}$ & 0.18$\pm$0.02 (0.18$\pm$0.02) & 0.45$\pm$0.01 (0.45$\pm$0.01) & 0.31$\pm$0.02 (0.31$\pm$0.02) & 0.06$\pm$0.01 (0.06$\pm$0.01) \\
    $f_{\rm star}$       & 0.02$\pm$0.00 (0.02$\pm$0.00) & 0.23$\pm$0.02 (0.20$\pm$0.01) & 0.52$\pm$0.01 (0.50$\pm$0.01) & 0.24$\pm$0.02 (0.28$\pm$0.02) \\
    $f_{\rm hot\ gas}$   & 0.00$\pm$0.00 (0.00$\pm$0.00) & 0.07$\pm$0.01 (0.00$\pm$0.00) & 0.35$\pm$0.02 (0.04$\pm$0.03) & 0.58$\pm$0.03 (0.96$\pm$0.03) \\
    $f_{\rm WHIM}$       & 0.02$\pm$0.00 (0.01$\pm$0.00) & 0.24$\pm$0.01 (0.18$\pm$0.01) & 0.58$\pm$0.01 (0.53$\pm$0.01) & 0.23$\pm$0.01 (0.28$\pm$0.02) \\
    $f_{\rm cold\ gas}$  & 0.20$\pm$0.02 (0.19$\pm$0.02) & 0.47$\pm$0.01 (0.47$\pm$0.01) & 0.28$\pm$0.02 (0.30$\pm$0.02) & 0.04$\pm$0.01 (0.05$\pm$0.01) \\
    \hline
  \end{tabular}
\end{table*}

We now investigate how the gas phase diagram behaves in different LSE (shown by columns) and at different redshifts (shown by rows) in Fig.~\ref{fig:gas_rte}. Note that the colour is normalised to the total gas mass to be consistent with Fig.~\ref{fig:gas_rt}. Again, the colour-map is coming from the AGN run whereas the contours are for the CSF. The gas is separated into different parts by black lines. We refer to Table~\ref{tab:env_bfrac} for detailed fractions in each LSE and redshift.

At $z = 0$, voids contain mainly cold gas; sheets are also dominated by cold gas, but there is a similar amount of WHIM; filaments have the largest portion of WHIM with a small fraction of cold gas; knots host the largest amount of hot gas with a small fraction of WHIM. As redshift increases, there is less and less WHIM and hot gas in the void as indicated by the colour map shrinking towards the lower temperature region. A similar conclusion can be drawn for the sheet environment for which the WHIM is also visibly decreasing, especially from $z=1.0$ to $z=2.1$. The WHIM dominates filament environments at $z = 0$, but its total amount also decreases with an increase of the cold gas fraction, especially from redshift $z = 1$ to $z = 2.1$. In the knot environment, the majority of gas remains at the warm/hot phase with a temperature of about 10$^7$ K until $z = 1.0$, whereas it is taken over by the cold gas at $z = 2.1$. This is because both the total WHIM and hot gas decrease very fast from redshift $z = 1$ to $z = 2.1$ (see Table~\ref{tab:cos_bfrac} for details).
Besides the difference at the lower temperature region between the two runs, which is already discussed in the previous subsection, the effect of AGN feedback is negligible.

\bigskip

We further detail the fractions of total gas, hot gas, WHIM, cold gas and stars by separating them into different LSE in Table~\ref{tab:env_bfrac}. For each gas component and stars, the fraction is normalised to their total mass in that baryon component, i.e. the sum of the fractions in the four LSE is equal to unity. At $z = 0$, it is clear that the majority of the gas can be found in sheet and filament environments. The hot gas is mainly divided between filament and knot environments. About half of the WHIM gas is located in filaments, followed by sheets ($\sim$30 per cent). About half of the cold gas is in sheets, the other half is almost equally assigned to filaments and voids. Most stars are in filaments (about 44 per cent), sheets (33 per cent) and knots (15 per cent). 

From low to high redshift, more and more (total) gas is found in sheets. This is the essence of structure formation: gas flows from voids to sheets then filaments and finally nodes. Therefore, towards lower redshifts, more and more gas flow into filaments and nodes. The stellar fraction increases mainly in filaments when moving backwards in time, indicative of star formation primarily taking place in filaments at redshifts for which the star formation rate peaks. The hot gas tends to shift from filament towards knot where feedback is prominent. The fractions of WHIM seem stable over redshift with always over half of them in filament. Similar to the WHIM, the fractions for cold gas are also independent of redshift with about half of them in sheets.

At last, there are some pronounced differences between the two hydrodynamical simulations. While they overall give similar fractions, the hot gas fractions at high redshift differ substantially. There is noticeable difference at $z = 2.1$, where 96 (58) per cent of hot gas is in knots from the CSF (AGN) run. This is again due to the powerful AGN feedback, which can heat up the gas and push it outside of the halos.

\section{Discussion and Conclusions}
\label{concl}

Using a series of cosmological simulations, which include baryonic processes with different levels of complexity, we study the distribution of matter in the large-scale environments (LSE) at four distinct redshifts. These sets of simulations allow us to explore how baryons affect the classification of the LSE: 1) a dark-matter-only run as a gauge, 2) a simulation that includes gas cooling, star formation and supernova feedback, and 3) a run that additionally models AGN feedback. To analyse all these simulations we employ the velocity shear tensor ({\sc Vweb}) based cosmic web classification on a regular grid of co-moving spacing $\sim 1.6 \Mpc$. Each cell is assigned a tag `knot', `filament', `sheet', or `void' depending on how many eigenvalues are above a certain threshold $\lambda_{th} = 0.1$, which serves as the web classifier.

In this paper, we extend the analysis presented in \citetalias{Cui2018} to higher redshifts $z>0$  and quantify the detailed baryon distributions. Our main results are summarised as follows
\begin{enumerate}
 \item In agreement with the result in \citetalias{Cui2018}, the baryon models have negligible effect on the LSE classification. Actually, the difference is smaller at redshifts $z > 0$. 
 \item Using the gas component alone to classify the LSE gives consistent results to the ones obtained accounting for all the matter. Again, the difference tends to be smaller at higher redshift. This (unsurprisingly) extends our initial finding that gas web is an unbiased tracer of total matter out to high redshift, especially for the filament environment. 
 \item The gas density PDFs indicate that knots host the densest gas and the mean gas density keeps dropping from filaments to voids. This is independent of the redshift evolution of the total gas density PDFs.
 \item By separating the gas mainly in temperature, we find about 40 per cent of gas is WHIM. The WHIM mass fraction drops with redshift, especially from $z = 1$ (29 per cent) to $z = 2.1$ (10 per cent).
 \item The detailed gas phase diagram shows the WHIM occupies the largest amount of gas in filaments. By separating the whole WHIM mass into the four LSE, about a half of the WHIM is in filaments and this fraction seems unchanged over redshift.
 \item We do not find significant difference between the two hydrodynamical runs for these fractions, besides the hot gas fraction in the knot environment at high redshift.
%  \item The cosmological stellar mass fraction is lower than observation in both hydrodynamical simulations implying that the theoretical models with AGN feedback may still be problematic. 
\end{enumerate}

We have shown that cold gas and WHIM share a similar fraction of gas mass at $z = 0$. However, most cold gas is located in the sheet environment while half of the WHIM mass is in the filament \citep[see also][]{Martizzi2018}. Therefore, we predict that the filament environment is the most promising place to look for WHIM and solving the missing baryon problem. However, because the whole WHIM mass fraction drops with increasing redshift, only $\sim 10$ per cent of total baryonic mass at $z=2.1$, the signal coming from the WHIM \citep[mostly in the soft X-ray band with the emission lines from highly ionised oxygen atoms such as OVII and OVIII, see ][for example]{Fang2005, Bertone2008, Bertone2010, vandeVoort2013}, should be much stronger at lower redshift. The largest fraction of gas at high redshift is in cold phase, which should be detected by SKA.

Our findings are generally in agreement with \cite{Martizzi2018}(see also \cite{Nevalainen2015} for a similar result on the WHIM fractions in filaments from hydro-simulation and \cite{Eckert2015} for the results from observations), who applied the tidal tensor method to the IllustrisTNG simulations. For example, the filament volume fractions are around 20 per cent with mass fractions of 40 per cent; the total hot gas mass fractions are 7.5\% versus 4.6\% at $z=0$, 1.6\% versus 1.3\% at $z=1.0$ and 0.24\% versus 0.2\% at $z\sim 2$; the total WIHM mass fractions are 49.9\% versus 41.3\%, 35\% versus 28.7\% at $z=1.0$ and 18.5\% versus 10.8\% at $z\sim 2$; the WHIM mass fractions to the total gas mass in filaments, sheets and knots are 0.27, 0.17, 0.03 in their study, whereas we obtain, respectively, 0.21, 0.12 and 0.06.

Although our simulation resolution is not very high, we believe that our results are almost unaffected by that. This is because our analysis mainly focuses on larger scales. Finer resolution may provide more correct figures at smaller scale. However, the overall baryon distributions, for example baryonic fractions within our mesh cell, are more model dependent than resolution dependent. As \cite{vanDaalen2011} (see their figure A2) have shown, the effect of resolution on the power spectrum is within 1 per cent at $k \approx 5 ~ h{\rm/Mpc}$, which is the scale corresponding to our cell size.

\section*{Acknowledgements}

We thank the referee for their thorough and thoughtful review of our paper.
The authors would like to thank Giuseppe Murante and Stefano Borgani for preparing these simulations, which have been carried out at the CINECA supercomputing Centre in Bologna, with CPU time assigned through ISCRA proposals and with the support from the PRIN-INAF12 grant 'The Universe in a Box: Multi-scale Simulations of Cosmic Structures', the PRINMIUR 01278X4FL grant 'Evolution of Cosmic Baryons', the INDARK INFN grant and 'Consorzio per la Fisica di Trieste'.

This work has made extensive use of the Python packages --- Ipython with its Jupyter notebook \citep{ipython}, NumPy \citep{NumPy} and SciPy \citep{Scipya,Scipyb}. All the figures in this paper are plotted using the python matplotlib package \citep{Matplotlib}. This research has made use of NASA's Astrophysics Data System and the arXiv preprint server. 

WC, AK, GY and RM are supported by the {\it Ministerio de Econom\'ia y Competitividad} and the {\it Fondo Europeo de Desarrollo Regional} (MINECO/FEDER, UE) in Spain through grant AYA2015-63810-P. WC further acknowledges the supported by the European Research Council under grant number 670193. AK further acknowledges funding through the Spanish Red Consolider MultiDark FPA2017-90566-REDC and thanks Blumfeld for l'etat et moi.
NIL acknowledges financial support of the Project IDEXLYON at the University of Lyon under the Investments for the Future Program (ANR-16-IDEX-0005).
XY is supported by the National Science Foundation of China (NSFC, grant Nos. 11833005, 11621303).
WC wishes to acknowledge support by the National Key R\&D Program of China (grant No. 2018YFA0404502) and by the NSFC (grant No. 11821303).
SP is ``Juan de la Cierva'' fellow (ref. IJCI-2015-26656) of the {\it Spanish Ministerio de Econom{\'i}a y Competitividad} (MINECO) and acknowledges additional support by the MINECO through the grant AYA2016-77237-C3-3-P.
HW acknowledges the support from the NSFC under grant No. 11733004 and 11421303.

\bigskip

%*****************************************************************************
\bibliographystyle{mnras}
\bibliography{bibliography}
\bsp

\label{lastpage}
\end{document}